\documentclass[prd,nofootinbib,floats,superscriptaddress,eqsecnum,tightenlines,11pt]{revtex4}

\usepackage{hyperref}
\usepackage{graphicx}
\usepackage{amsmath,amssymb,amsfonts,amsthm,latexsym,stmaryrd}
\usepackage{marginnote}
\usepackage{color}
\usepackage{soul}

\newcommand{\beq}{\begin{equation}}
\newcommand{\eeq}{\end{equation}}

\newcommand{\U}{u}
\newcommand{\ta}{a}
\newcommand{\tb}{b}
\newcommand{\ptwo}{\sigma}

\begin{document}

\title{Neutron stars in degenerate higher-order scalar-tensor theories}
\author{Hamza Boumaza}
\affiliation{Laboratoire de Physique des Particules et Physique Statistique (LPPPS),\\
 Ecole Normale Supérieure-Kouba, B.P. 92, Vieux Kouba, 16050 Algers, Algeria}
\affiliation{Laboratory of Theoretical Physics and Department of Physics,\\
Faculty of Exact and Computer Sciences, University Mohamed Seddik Ben Yahia, BP 98, Ouled Aissa, Jijel 18000, Algeria}
\author{David  Langlois}
\affiliation{Université Paris Cité, CNRS, AstroParticule et Cosmologie, F-75013 Paris, France}
\date{\today}

\begin{abstract}
We study neutron star configurations in a simple  shift-symmetric subfamily of degenerate higher-order scalar-tensor  (DHOST) theories, 
 whose deviations from General Relativity (GR) are  characterized by a single parameter. 
 We compute the radial profiles of neutron stars in these  theories of modified gravity, using several realistic equations of state for the neutron star matter.  We find neutron stars with masses and radii significantly larger than their GR counterparts.
 We then consider slowly-rotating solutions and determine the relation between the dimensionless moment of inertia and the compactness, relation that has the property to be almost insensitive to the equation of state.
\end{abstract}

\maketitle

\section{Introduction}

Interesting constraints on modified gravity can be obtained  by confronting astrophysical predictions of modified gravity models with  observations, including precious data from  gravitational waves \cite{abbott2016gw150914,abbott2018gw170817,abbott2018prospects,chornock2017electromagnetic}.
This is in particular the case  for astrophysical objects  with strong gravity  where
 one could expect significant deviations from General Relativity (GR). Since black holes in modified gravity are often indistinguishable from their GR counterparts, as a consequence of generalised no-hair theorems \cite{Hui:2012qt}, albeit not always (see e.g.  \cite{Babichev:2017guv}), it is natural to turn to the next astrophysical objects where strong gravity is present: neutron stars.

 In the present work, we study neutron stars in the context of   Degenerate Higher-Order Scalar-Tensor (DHOST) theories. These  theories, introduced in \cite{Langlois:2015cwa,Langlois:2015skt}   (and further explored in \cite{BenAchour:2016cay,Crisostomi:2016czh,BenAchour:2016fzp}) represent  the largest family of scalar-tensor theories propagating a single scalar degree of freedom    
(see \cite{Langlois:2018dxi} for a review).  Given a  model of neutron star, it is useful  to compute  several  global properties that in principle can be extracted from present or future data, such as  the mass, radius and moment of inertia. The comparison of these predictions with observed neutron stars then provides constraints on the equation of state of neutron star matter, which  so far remains largely unknown, as well as  on the underlying gravity model. 

The behaviour of high density nuclear matter in the  core of neutron stars remains an open question, both theoretically and observationally. There exists in the literature a broad spectrum of equations of state, which lead to different predictions for the bulk properties of neutron stars, such as the maximal mass or the  mass-radius relation (see e.g. \cite{Lattimer:2015nhk}). This uncertainty on the equation of state makes the investigation of gravity within neutron stars more difficult since there could be degeneracies between potential deviations from GR and variations of  the equation of state. 

Fortunately, it has been shown in the context of GR that one can 
find  relations between observable quantities, for example between  the moment of inertia and the compactness, that are relatively insensitive to the choice of equation of state \cite{Lattimer:2004nj,Breu:2016ufb}.  One can expect that modified gravity  models  lead to  similar but distinct relations that could help constrain these models, independently of the equation of state. One of the goals of this paper is to determine the new relations  between the moment of inertia and compactness in the gravity models we explore, which requires to consider slowly-rotating neutron stars.

Neutron stars have already been studied for a few particular subfamilies of DHOST theories  in several works  \cite{Babichev:2016jom,Minamitsuji:2016hkk,Cisterna:2016vdx,Sakstein:2016oel,Kobayashi:2018xvr,Chagoya:2018lmv,Boumaza:2021lnp,Boumaza:2021hzr}.
Some of the works on neutron stars in DHOST theories or subfamilies have also considered slowly-rotating solutions: in Horndeski theories \cite{Cisterna:2016vdx} or in Beyond Horndeski theories \cite{Sakstein:2016oel}.

The outline of our paper is the following. After introducing, in the next section, the DHOST theories and the models we consider in this work, we then derive, in section \ref{section:eom},  the main equations for a static and spherically symmetric configuration.  In section \ref{section:NS}, we solve the system for five realistic equations of state, and obtain a continuum of neutron star solutions parametrised by their central energy density. In the subsequent section, we extend our discussion to slowly-rotating neutron starts and determine the  quantitative  relation between compactness and moment of inertia. We finally give some conclusions and perspectives in the final section. We have also added two appendices, where some technical details are presented.

\section{Gravity models}

The most general scalar-tensor theories propagating a single scalar degree of freedom are known as DHOST theories. They encompass the traditional scalar-tensor theories as well as Horndeski theories and, in contrast to the latter, are closed with respect to the most general disformal transformations of the metric, which are field redefinitions mixing the metric and scalar field. \\
In the present work, we restrict ourselves to quadratic\footnote{This means that the Lagrangian is quadratic in second derivatives of the scalar field $\nabla_\mu\!\nabla_\nu\varphi$.} shift-symmetric DHOST theories.
The corresponding total action reads
\begin{equation}
S=\int d^4x\sqrt{-g}\left(F_{0}+F_1\nabla^{\mu}\nabla_{\mu}\varphi+F\,R+\sum_{i=1}^{i=5}A_{i}L_{i}+L_{\rm m}\right)\label{Stot}
\end{equation}
where the functions $A_{i},\,F,\,F_1$ and $F_{0}$ depend only on  $X\equiv\nabla^{\mu}\varphi\nabla_{\mu}\varphi$, $R$ is the Ricci Scalar and the five elementary Lagrangians $L_{i}$ that are quadratic in second order derivatives of the scalar field are given by 
\begin{eqnarray*}
L_{1}=\nabla^{\nu}\nabla^{\mu}\varphi\nabla_{\nu}\nabla_{\mu}\varphi & L_{2}=\left(\nabla^{\mu}\nabla_{\mu}\varphi\right)^{2} & L_{3}=\nabla^{\alpha}\nabla_{\alpha}\varphi\nabla_{\nu}\nabla_{\mu}\varphi\nabla^{\nu}\varphi\nabla^{\mu}\varphi\\
 & L_{4}=\nabla^{\nu}\varphi\nabla_{\nu}\nabla_{\mu}\varphi\nabla^{\mu}\nabla^{\alpha}\varphi\nabla_{\alpha}\varphi\,, \qquad
 & L_{5}=\left(\nabla^{\nu}\varphi\nabla_{\nu}\nabla_{\mu}\varphi\nabla^{\mu}\varphi\right)^{2}\,.
\end{eqnarray*}
We have also added
$L_m$, the Lagrangian describing matter, which is assumed to be minimally coupled to the metric $g_{\mu\nu}$.

The functions $F$, $F_1$ and $F_{0}$ are arbitrary. Among the functions $A_i$, two of them can be chosen arbitrarily, for example $A_1$ and $A_3$, but the three other ones must be  related to the first two ones in order the theory to be degenerate, and thus to contain a single scalar degree of freedom. These relations, which are direct  consequence of the degeneracy conditions, are given by \cite{Langlois:2015cwa}
\begin{eqnarray*}
A_{2} & = & -A_{1}\\
A_{4} & = & \frac{1}{8\left(F-X A_{1}\right)^2}\Bigl(A_{1}A_{3}\left(12XF-16X^{2}F_{X}\right)+4A_{1}^{2}\left(16XF_{X}+3F\right)-16A_{1}\left(4XF_{X}+3F\right)F_{X}\\
 &  & -16XA_{1}^{3}+8A_{3}F\left(XF_{X}-F\right)-X^{2}A_{3}^{2}F+48FF_{X}^{2}\Bigr)\\
A_{5} & = & \frac{1}{8\left(F-XA_{1}\right)^2}\Bigl(4F_{X}-2A_{1}+XA_{3}\Bigr)\Bigl(4A_{1}F_{X}-3XA_{1}A_{3}-2A_{1}^{2}+4 F A_{3}\Bigr),
\end{eqnarray*}

For simplicity, in the present work we restrict our study to the cases where 
\beq
\label{A1_A2}
A_{1}=-A_{2}=0\,,
\eeq
and 
\beq
A_{3}=-A_{4}=-4\frac{F_X}{X}\,,
\eeq
which also implies $A_5=0$. 
Note that condition (\ref{A1_A2}) implies  the strict equality  between the speeds of light and gravitational waves \cite{Langlois:2017dyl}. 
We stress that  this constraint, usually invoked in the wake of   GW170817, is not necessary if one does not seek to account for dark energy, as a much larger region of the parameter space becomes available.

Moreover,  we also assume $F_0=0$ and $F_1=0$, so that we effectively  work with the action
\begin{equation}
S_{\rm grav}=\int d^{4}x\, \sqrt{-g}\left(F\,R-4\frac{F_X}{X}(L_3-L_4)+L_{\rm m}\right)\,,
\label{S}
\end{equation}
which depends on a single function $F(X)$.
One can notice that the above models belong to the 'Beyond Horndeski' subfamily of DHOST theories, introduced in  \cite{Gleyzes:2014dya,Gleyzes:2014qga}.

\section{Equations of motion}
\label{section:eom}
We now wish to study a relativistic star in the theories of modified gravity (\ref{S}). Let us start with the simpler case of nonrotating stars, before considering  slowly-rotating stars in  section \ref{section:rotating}. For a non-rotating star, the metric is static and  spherically symmetric, i.e. of the form
\begin{equation}
ds^{2}=-f(r)dt^{2}+h(r)dr^2+r^2\left(d\theta^2+\sin^2\!\theta\,  d{\phi}^2\right).
\label{ds}
\end{equation}
We assume  that the scalar field  takes the form 
\beq
\label{phi}
\varphi(t,r)=q\;t+\psi(r)\,,
\eeq
where a linear time dependence is allowed for shift-symmetric theories  since the gradient of $\varphi$ is time-independent \cite{Babichev:2013cya}.

 Substituting the above metric (\ref{ds}) and scalar field (\ref{phi}) into the action (\ref{S}) we get, after integrating by parts, the expression
\begin{equation}
S=\int d^{4} x \, r^2\sqrt{f h}\left(\frac{2 F h'}{h^2 r}+\frac{2 F (h-1)}{h r^2}-\frac{4 F_X X' \left(f X+q^2\right)}{f h r X}+L_m\right)\,,
\label{S2}
\end{equation}
with
\beq
\label{X}
X=\psi'^2/h-q^2/f\,.
\eeq
We do not need to specify the matter  Lagrangian  $L_m$  but only its variations with respect 
 to the metric $g_{\mu\nu}$. In practice, the matter is modelled  by a perfect fluid with energy density $\rho$, pressure $P$ and 4-velocity $u^\mu$. The  energy-momentum is thus given by 
\beq
T^{\mu\nu}=\frac{2}{\sqrt{-g}}\frac{\delta (\sqrt{-g}\,  L_m)}{\delta g_{\mu\nu}}=(\rho+P)u^\mu u^\nu + Pg^{\mu\nu}\,.
\eeq
The conservation of the energy-momentum tensor $\nabla_\mu T^{\mu\nu}=0$ then implies
\begin{eqnarray}
P'=-\frac{f'(P+\rho)}{2f}\,.
\label{e4}
\end{eqnarray}
For an equation of state $P=P(\rho)$, which is appropriate for the neutron star interior, we also have the relation
\beq
P'=c_m^2 \, \rho'\,,
\label{sound}
\eeq
where $c_m$ denotes the sound speed. 

\medskip
 By varying the action (\ref{S2}) with respect to $f$ and $h$, we obtain the time and radial equations of motion which read, respectively,
\begin{eqnarray}
&&h' \left(\frac{F_{\text{X}} \left(-4 f^2 \left(\psi '\right)^4+8 f h q^2 \left(\psi '\right)^2+4 h^2 q^4\right)}{f h^2 q^2-f^2 h \left(\psi '\right)^2}+2 F\right)-\frac{8 F_{\text{X}} \left(f \left(\psi '\right)^3-2 h q^2 \psi '\right)}{f \left(\psi '\right)^2-h q^2} \psi '' \nonumber\\
&&+\frac{4 h q^2 F_{\text{X}} \left(f (h+1) \left(\psi '\right)^2-(h-1) h q^2\right)}{f r \left(f \left(\psi '\right)^2-h q^2\right)}+\frac{2 F (h-1) h}{r}-h^2 r \rho  =0,\label{eh}\\
&&f' \left(\frac{F_{\text{X}} \left(-4 f^2 \left(\psi '\right)^4+8 f h q^2 \left(\psi '\right)^2+4 h^2 q^4\right)}{f^2 h^2 q^2-f^3 h \left(\psi '\right)^2}+\frac{2 F}{f}\right)+\frac{8 q^2 F_{\text{X}} \psi ' \psi ''}{h q^2-f \left(\psi '\right)^2}\nonumber\\
&&+\frac{F_{\text{X}} \left(4 (h-1) h q^2 \left(\psi '\right)^2-4 f (h+1) \left(\psi '\right)^4\right)}{h r \left(h q^2-f \left(\psi '\right)^2\right)}+\frac{F (2-2 h)}{r}-h r P=0.
\label{ef}
\end{eqnarray}
In GR, where $F=\kappa/2=c^4/(16\pi G)$, $G$ being Newton's constant, these equations reduce, after division by $2F=\kappa$,  to 
\begin{eqnarray}
&&h' +\frac{(h-1) h}{ r}-\frac{h^2 r  \rho }{\kappa  }=0,\label{eh_GR}\\
&& \frac{  f'}{f}- \frac{(h-1) }{ r}-\frac{h r P}{\kappa }=0,\label{ef_GR}
\end{eqnarray}

Finally, by varying  the action \eqref{S2} with respect to $\psi$, we obtain the scalar field equation of motion. In the shift-symmetric case, this is related to the conservation of a four-dimensional current, $\nabla_\mu J^\mu=0$, which reduces, due to the symmetries of the configuration, to 
\begin{eqnarray}
\frac{d}{dr}\left[J\right]=0,
\end{eqnarray}
with
\begin{eqnarray}
J=\frac{2 \sqrt{f X+q^2}}{f^{5/2} h^{5/2} X}\left\{-2 F_X \left[h r f' \left(f X-q^2\right)+f \left(f h (h+1) X-q^2 r h'+2 h q^2\right)\right]\right\},
\end{eqnarray}
which implies
\beq
J=0\,.
\eeq
This is an equation for $X$, or equivalently for 
$\psi'^2$, which can be solved explicitly once we specify the function $F(X)$. 
A simple form for $F$ is 
\beq
F=\frac{\kappa}{2}+\ptwo X\,,
\eeq
which can be seen as the first two terms in a Taylor expansion with respect to $X$.

With this ansatz, the equation $J=0$ gives a quadratic equation for $\psi'^2$. One of its  solutions can be expressed, using (\ref{eh}) and (\ref{ef}), in the form\footnote{We have retained the solution that behaves as in GR asymptotically.}
\begin{eqnarray}
\psi'^2 =\Delta_1-\Delta_2,\label{ex}
\end{eqnarray}
where
\begin{eqnarray}
\Delta_1 &=& \frac{h \left[f \left(2 \kappa + r^2 P\right)+8 p \right]-4 p}{8 f}\\
\Delta_2^2 &=& \Delta_1^2-\frac{h p \left[f \left(h \left(2 \kappa +2 P r^2+ \rho  r^2\right)-2 \kappa \right)+4 p
   \left(h -1\right)\right]}{4 f^2} 
\end{eqnarray}
In the above equations,  we have renormalized the scalar field  and the parameter  $\ptwo$  as follows: 
\beq
\psi_{\rm new}=\frac{p^{1/2}}{q} \psi_{\rm old}\,, \qquad p\equiv  q^2 \ptwo\,.
\eeq
With these redefinitions, the constant  $q$ disappears from the equations and our modified gravity models are characterised by the single parameter $p$.
In comparison with  the coefficients characterizing the deviations from standard gravity  defined in \cite{Langlois:2017dyl},
we find in our case
\begin{eqnarray}
\Xi_1 =-2\, \Xi_2 = \frac{2p}{2p+\kappa}.
\end{eqnarray}

\medskip
Outside the star, where $P=0$ and $\rho=0$, one can check that  the metric  and scalar field equations  are satisfied by inserting  the usual Schwarzschild metric functions
\begin{eqnarray}
f=h^{-1}=1-2\frac{M}{r},\label{f}
\end{eqnarray}
where the integration constant  $M$ corresponds to  the mass of the neutron star, as well as the scalar field profile
\begin{eqnarray}
\psi'=\frac{\sqrt{2 p M  r}}{r-2 M}.
\label{psi_ext}
\end{eqnarray}
This shows in particular that the constant $p$, i.e. $\ptwo$,  must be positive.  Substituting (\ref{f}) and (\ref{psi_ext}) into (\ref{X}), one finds that $X=-q^2$ outside the star.

In summary, as far as the exterior of the neutron star is concerned, the above vacuum solution  is indistinguishable from GR. We turn to the interior of the star in the next section.

\section{Neutron star profiles}
\label{section:NS}
In this section, we compute the radial profile for various relativistic stars, depending on their central density and equation of state, by integrating numerically the equations of motion obtained in the previous section.  To do so, it is convenient to rewrite the equations of motion in the matricial  form
\beq
\label{system}
A\, \frac{dY}{dr}= B\,,\qquad Y=(f, h, \psi')^T\,,
\eeq
where the first line corresponds to (\ref{ef}) and the second line to (\ref{eh}). The last line is obtained from the radial derivative of Eq.(\ref{ex}), where the derivatives $P'$ and $\rho'$ can  be eliminated in favour of $f$, $f'$, $\rho$, $P$ and $c_m^2$ by using (\ref{e4}) and (\ref{sound}). The coefficients of the $3\times 3$ matrix $A$ and of the column matrix $B$ are given explicitly in the appendix. 
By multiplying (\ref{system}) by the inverse matrix $A^{-1}$, so that $Y'=A^{-1} B$, we obtain a first-order system of equations for the functions $f$, $h$ and $\psi'$, which can be  integrated numerically.

We have considered several  realistic equations of state discussed in the literature. Following  \cite{Haensel:2004nu,Potekhin:2013qqa}, they can be parametrised in the form
\begin{eqnarray}
\log_{10}\left(\frac{P}{{\rm g\;cm}^{-3}}\right) &=& \frac{\tb_1+\tb_2 \xi +\tb_3 \xi ^3}{1+\tb_4 \xi}\U\left[\tb_5 (\xi -\tb_6)\right]+(\tb_7+\tb_8 \xi)\U\left[\tb_9 (\tb_{10}-\xi )\right]+(\tb_{11}+\tb_{12} \xi)\U\left[\tb_{13} (\tb_{14}-\xi )\right]\nonumber\\
& & +(\tb_{15}+\tb_{16} \xi )\U\left[\tb_{17} (\tb_{18}-\xi )\right]
 +\frac{\tb_{19}}{\tb^2_{20} (\tb_{21}-\xi )^2+1}+\frac{\tb_{22}}{\tb^2_{23} (\tb_{24}-\xi )^2+1},\label{EOS}
\end{eqnarray} 
with the function
\begin{eqnarray}
\U[x]=\frac{1}{e^x+1}\,,\qquad  \xi=\log_{10}({\rho}/g\;cm^{-3})\,.
\end{eqnarray}
Each equation of state is characterised by the values of the coefficients $\tb_i$. 
 For the  SLy and FPS equations of state, these coefficients are given by
 \beq
 \tb_i=a_i^{\rm HP} \quad {\rm for }\quad 1\leq i \leq 18\,,\qquad  \tb_j = 0\quad {\rm for} \quad 19\leq j\leq 24\,,
 \eeq
 where the  $a_i^{\rm HP}$ denote the coefficients $a_i$ of \cite{Haensel:2004nu}.
For the BSk19, BSk20 and BSk21 equations of state,  the coefficients are
\beq
 \tb_i=a_i^{\rm PFCPG} \quad {\rm for }\quad 1\leq i \leq 9\,,\qquad \tb_{10}= a_6^{\rm PFCPG}\,, \qquad   \tb_j =a_{j-1}^{\rm PFCPG} \quad {\rm for} \quad 11\leq j\leq 24\,,
 \eeq
 where the  $a_i^{\rm PFCPG}$ correspond to  the coefficients $a_i$ of 
\cite{Potekhin:2013qqa}.

\begin{figure}[h]
\centering
\includegraphics[scale=0.7]{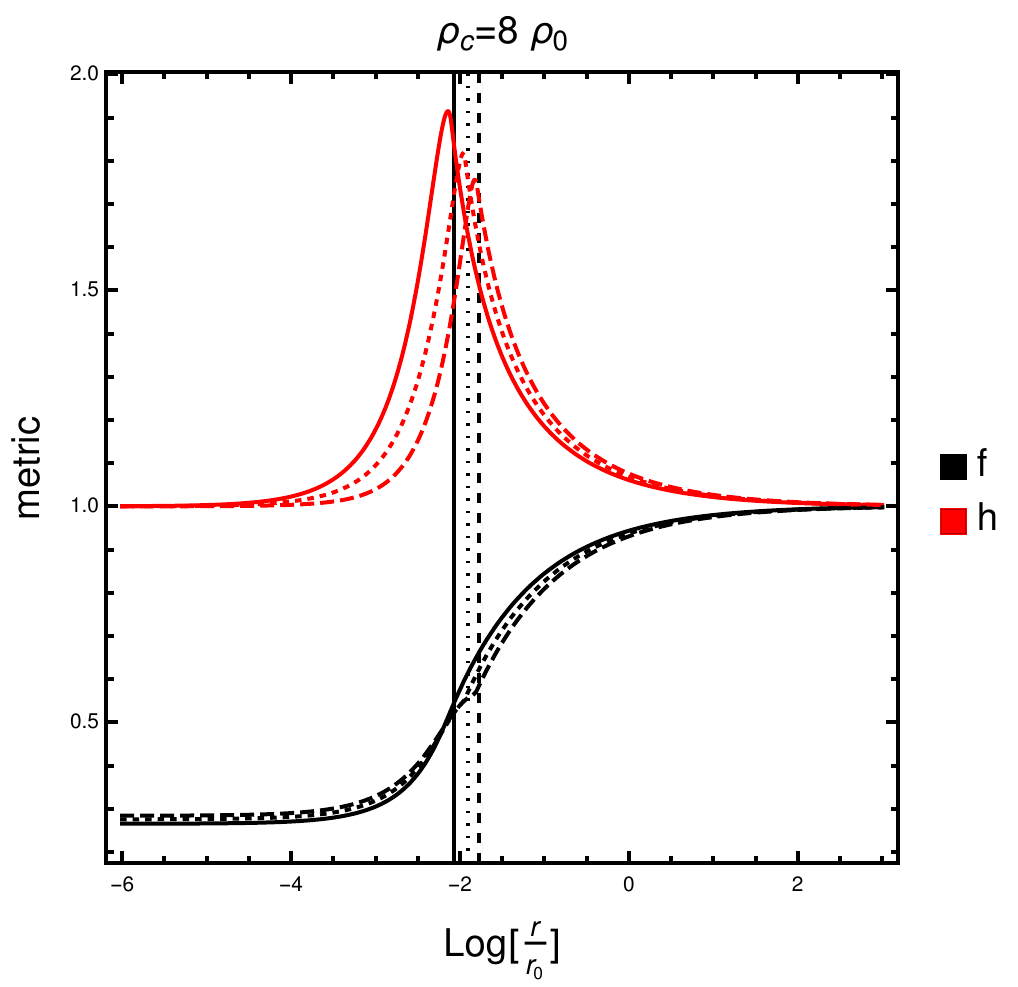}  
\includegraphics[scale=0.7]{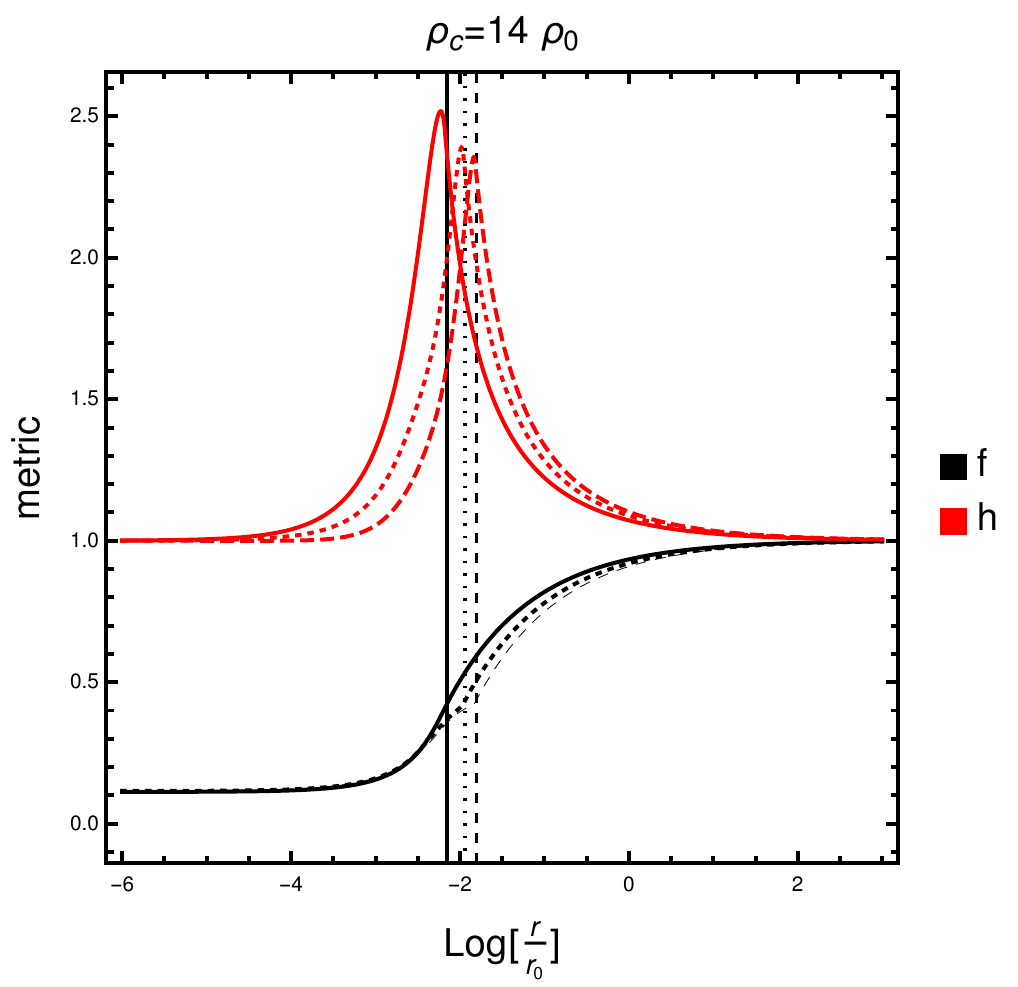} 
\caption{\small Radial profiles of $f$ and $h$, using the SLy equation of state,  in the cases $p=0$ (continuous line), $p=0.001$ (dotted line) and $p=0.002$ (dashed line), for two values of the central energy density $\rho_c$. The vertical bars correspond to the respective radii of the neutron stars.}
\label{f_h}
\end{figure}

\begin{figure}[h]
\centering
\includegraphics[scale=0.8]{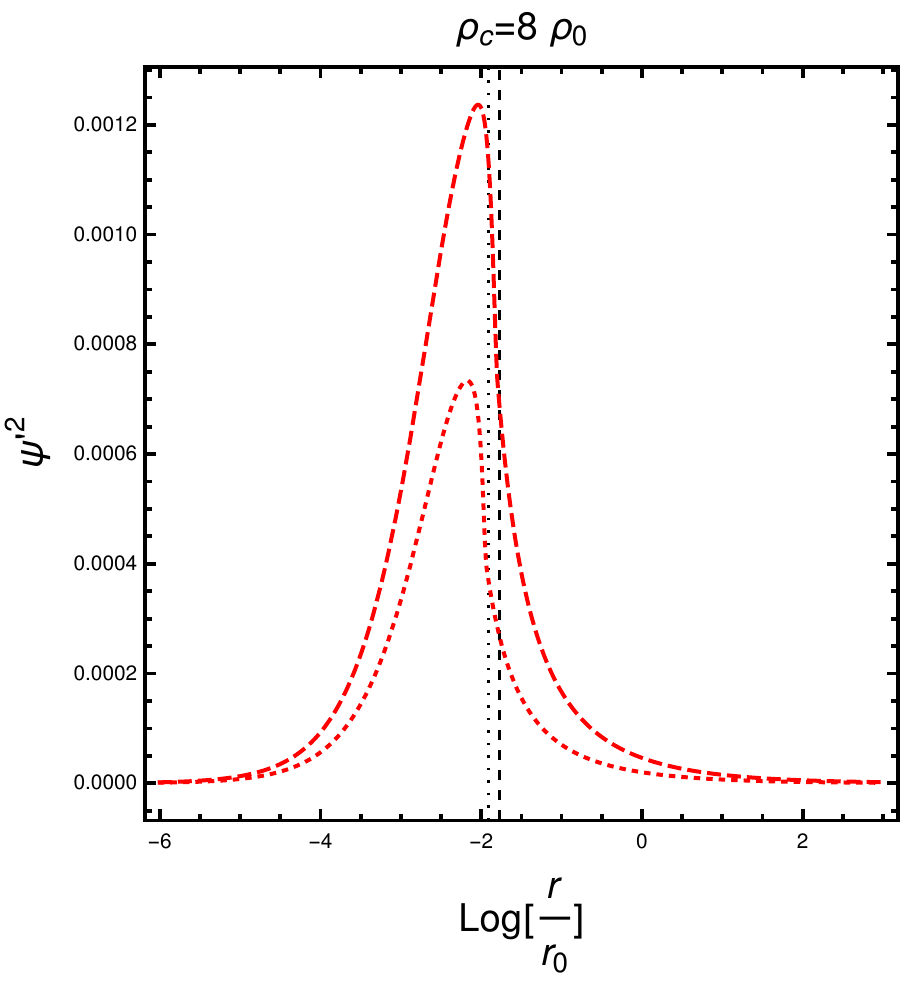}  
\includegraphics[scale=0.8]{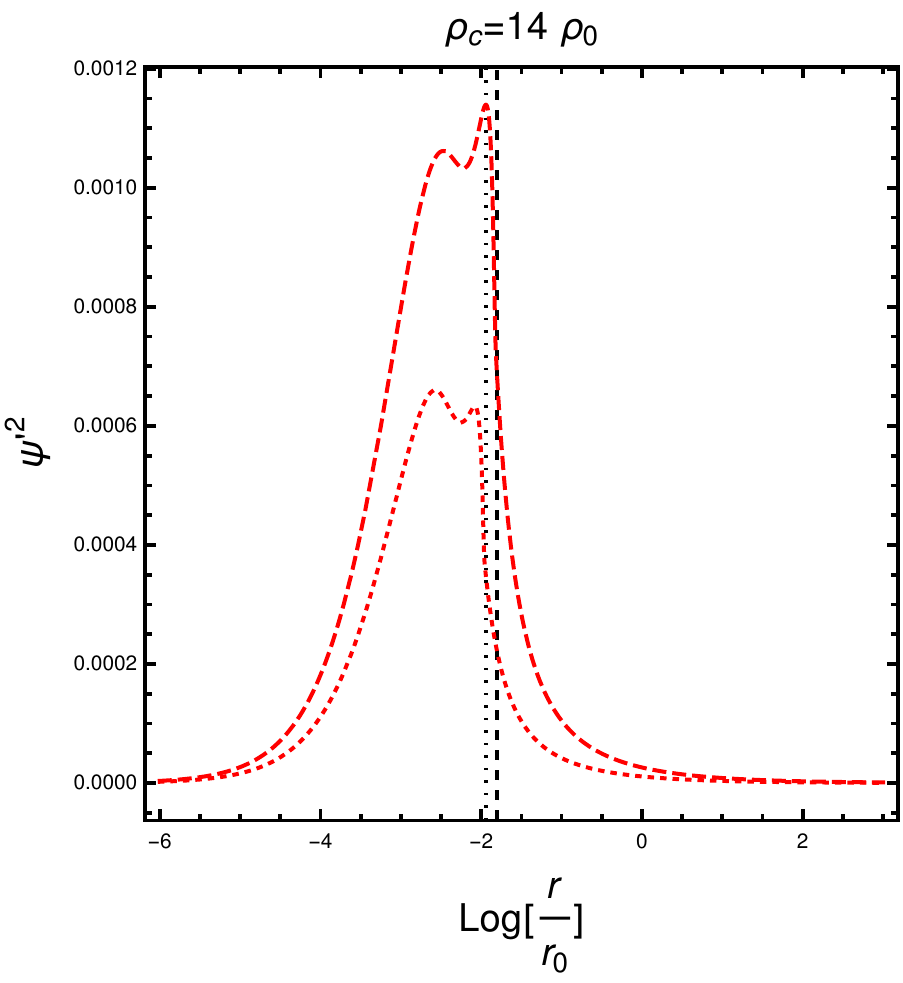}
\caption{\small Radial profile of $\psi'^2$ in the cases  $p=0.001$ (dotted line) and $p=0.002$ (dashed line), with the same parameters as in Fig.~\ref{f_h}.}
\label{fig_psi'}
\end{figure}

Given an equation of state and a gravity model characterised by the choice of the parameter $p$,  the radial profile of the star depends on the central energy density $\rho_c$. The other quantities at $r=0$ can be expressed in terms of $\rho_c$, as 
 discussed in the appendix.
For the integration, we  use the dimensionless variable $s=\ln(r/r_0)$, defined in terms of  the lengthscale
\beq
 r_0=\frac{c}{\sqrt{G \rho_0}}=89.664\, {\rm km}\,, \qquad \rho_0=m_{\rm n} n_0=1.6749\times 10^{14} {\rm g.cm}^{-3}\,,
 \eeq
 where $m_{\rm n}$ is the neutron mass and $n_0=0.1\; {\rm fm}^{-3}$ is the typical number density  in neutron stars.
Integrating the system of equations from $s=-\infty$ (in practice $s\approx -10$) to $s=\infty$ (in practice $s\approx 5$), we  obtain the radial  profiles of $f$, $h$ and $\psi'$, as illustrated for the first two quantities  in  Fig.~\ref{f_h}.

\begin{figure}[h]
\centering
\includegraphics[scale=0.7]{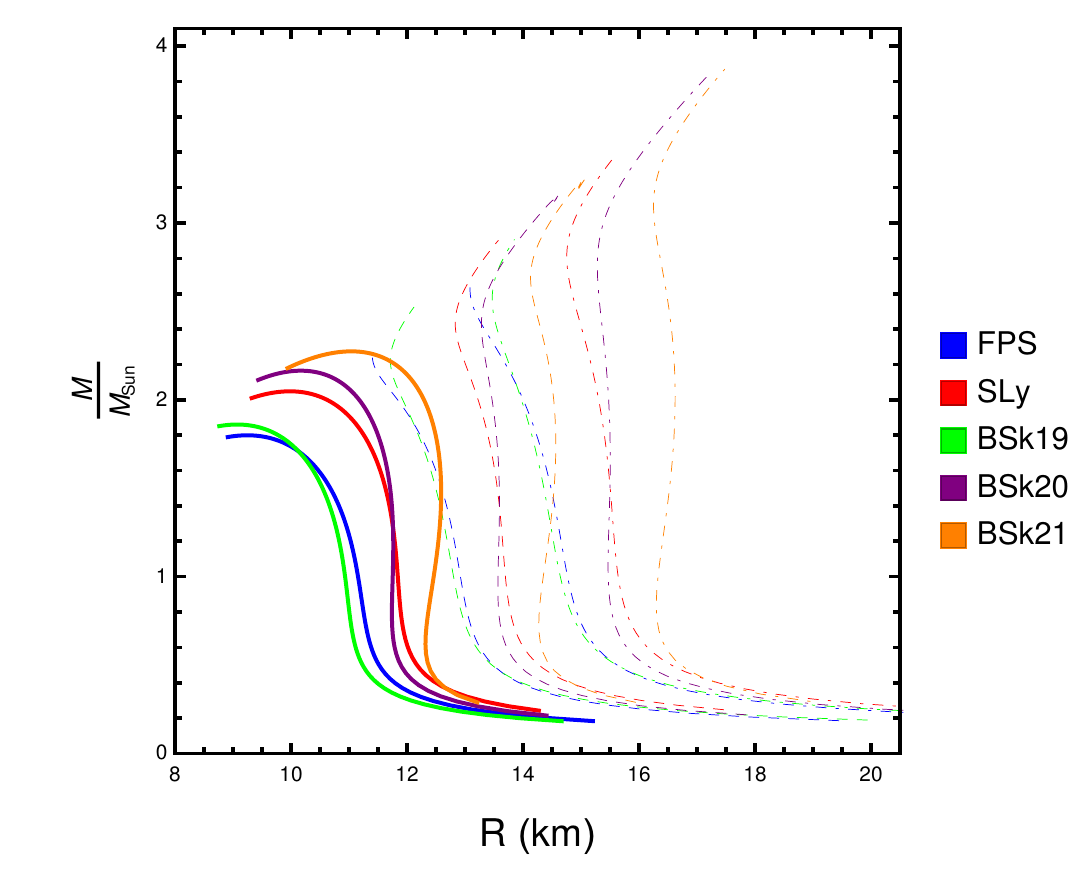}  
\caption{\small  Mass-radius relation for $p=0$ (continous lines), $p=10^{-3}$ (dashed lines), $p=2\times 10^{-3}$ (dot-dashed lines)  for various equations of state, distinguished by the colours.}
\label{mrs}
\end{figure}

\medskip

The radius of the star, denoted $R$,  is then determined by the condition  $P(R)=0$ at the surface of the star. After  the numerical resolution is achieved the mass of the star is also calculated,  using equation (\ref{f}). Our results are summarized  in Fig.\ref{mrs} for two values of the parameter $p$ and five different equations of state, varying the central density from $2\rho_0$ to $25\rho_0$. 

In fig. \ref{mrs}, we observe that the radius and the mass of the star become larger than those in GR if we increase the value of $p$. We also note observe that the maximum of the mass can be larger than $2 M_\odot$, where $M_\odot$ is the mass of the sun, for the FPS and BSk19
EoS. 
This is an interesting property in view of results such as  the observation the pulsar PSR J1614-2230 \cite{Demorest:2010bx}, or the mass of  the compact object measured  from the GW190814 event ($2.59 M_\odot$) \cite{abbott2020gw190814}.

\begin{figure}[h]
\centering
\includegraphics[scale=0.7]{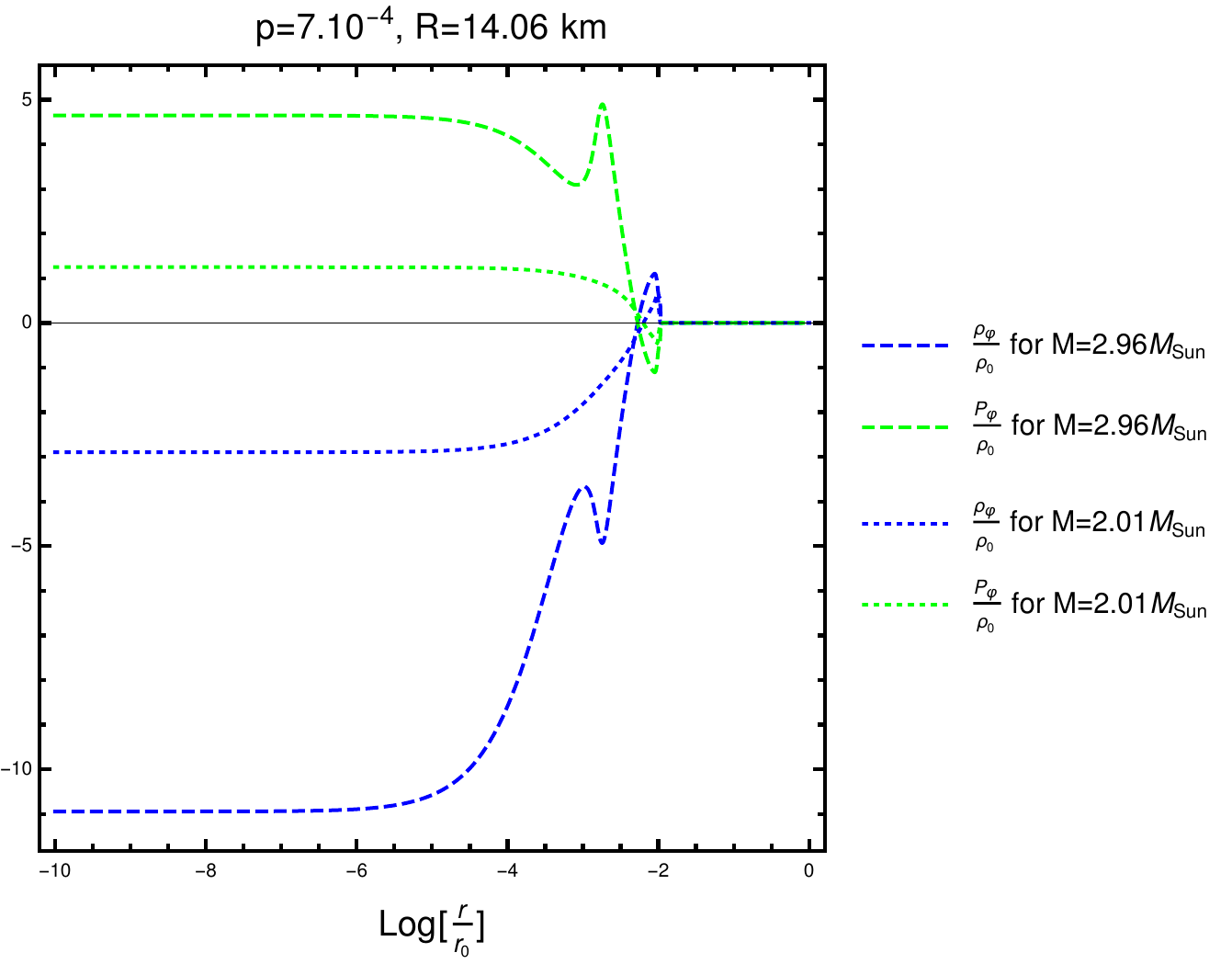}  
\caption{\small  Radial profiles of the effective energy density $\rho_\varphi$  and pressure $P_\varphi$  for two star configurations with the same radius 
$R=14.06 \, km$  but distinct masses, $M=2.01 M_\odot$ and $M=2.96 M_\odot$, corresponding respectively to the central energy densities $\rho_c \simeq 7.9  \rho_0$ and $\rho_c \simeq 21.3 \rho_0$.}
\label{rhoeff}
\end{figure}

In order to better understand how the neutron stars obtained in these modified gravity models can differ so much  from their GR counterparts, 
it is useful to introduce the effective energy density and effective pressure associated with the  scalar field. They can be defined simply by rewriting the equations (\ref{eh})-(\ref{ef}) in the GR form (\ref{eh_GR})-(\ref{ef_GR}) , up to a replacement of $\rho$ and $P$ by $\rho+\rho_{\varphi}$ and $P+P_\varphi$, respectively, so that
\begin{eqnarray}
\rho_\varphi &=&\kappa  \left(\frac{h'}{h^2 r}-\frac{1}{h r^2}+\frac{1}{r^2}\right)-\rho,\\
P_\varphi &=& \kappa  \left(\frac{f'}{f hr}+\frac{1}{h r^2}-\frac{1}{r^2}\right)-P\,.
\end{eqnarray}
As an example, we plot in Fig. \ref{rhoeff} the radial profiles of $\rho_\varphi $ and  $P_\varphi $ for two neutron stars with the same radius but very different masses, assuming the SLy equation of state and for $p=7.10^{-4}$. We find  that the effective scalar field energy density and  pressure are important, which explains why the deviation of the mass and  radius from their GR values can be considerable. Interestingly, the pressure and the energy density have opposite sign, with $P_\varphi$  positive from $r=0$ to $r\sim 10$ km and then negative till $r=R$ for both $M=2.96 M_\odot$ and $M=2.01 M_\odot$. Morever, the number of critical points  (defined as local extrema) in the profiles of $P_\varphi $ and $\rho_\varphi $ is different for the two neutron stars. 

\section{Slowly-rotating neutron stars}
\label{section:rotating}
We now extend our study to  slowly-rotating stars, following the method proposed by  Hartle and Thorne~\cite{Hartle:1967he,Hartle:1968si}.
In order to derive the equations for a slowly-rotating neutron star, we generalise  the static metric  \eqref{ds}  to the new metric 
\begin{eqnarray}
ds^{2}=-f(r)dt^{2}+h(r)dr^2+r^2\left(d\theta^2+\sin^2\!\theta\,  d{\phi}^2\right)
+2 w(r, \theta) r^2 \sin^2\!\theta\,  dt d\phi,\label{ds2}
\end{eqnarray}
where the last term, associated with the rotation, is assumed to be small, i.e. $w(r, \theta)\ll f$. Similarly, the perfect fluid in the star is now rotating, which can be described by the four-velocity
\begin{eqnarray}
u^\mu = \frac{1}{f^{1/2}}\{\;1\;,\;0\;,\;0\;,\;\Omega\; \}+{\cal O}\left(\Omega^2 \right)\,,
\end{eqnarray}
up to first order in the fluid angular velocity $\Omega=d\phi/dt=u^\phi/u^t$.

Substituting the new metric (\ref{ds2}) into the action \eqref{S} and expanding it in terms of $w$ up to quadratic order,  we obtain, after  integration by parts, the following quadratic action for the function $w(r,\theta)$:
\begin{eqnarray}
S^{(2)}=\int d^4 x \sin^3\theta \left(\frac{F r^4 }{2 \sqrt{f} \sqrt{h}} (\partial_r w)^2+\frac{F \sqrt{h} r^2 }{2 \sqrt{f}}(\partial_\theta w)^2+\frac{\sqrt{h} \rho  r^4 }{2 \sqrt{f}}w^2+L_m^{(2)}\right)\,.
\end{eqnarray} 
By varying $S^{(2)}$ with respect to $w$ and taking into account the matter action, we get
\begin{eqnarray}
&&\partial_{rr} w+ \left(\frac{4}{r}-\frac{f'}{2 f}-\frac{h'}{2 h}+\frac{F_X X'}{F}\right) \partial_{r}w+\frac{h}{r^2} \left(\partial_{\theta\theta} w+3 \cot\theta\, \partial_{\theta}w\right)\nonumber\\
&&-\frac{h   (P+\rho )}{F}(\Omega+w)=0\,.
\label{eq_omega}
\end{eqnarray}
One can solve this partial differential equation by looking for separable solutions.  Decomposing   $w$  as 
\begin{eqnarray}
w(r,\theta)= \sum_{l} w_l (r) \, P_{l}(\theta),
\label{omegalegndre}
\end{eqnarray}
where $P_{l}$  are the Legendre polynomials,  and substituting into  (\ref{eq_omega}),  one finds that the radial functions $w_l(r)$ must satisfy the ordinary differential equations
\begin{eqnarray}
&&\frac{(hf)^{\frac{1}{2}}}{r^4 F}\frac{d}{dr}\left[\frac{r^4 F}{(hf)^{\frac{1}{2}}}w_l'\right]+\frac{h ((l+1)l-2)}{r^2} w_l-\frac{h   (P+\rho )}{F}(\Omega+w_l)=0.\label{omega}
\label{eq_w_l}
\end{eqnarray}
Outside the star, as discussed at the end of section \ref{section:eom}, we know that  the metric functions $f$ and $h$ are the same as in Schwarzschild and $F(X)$ is constant, so that we recover the same equations as in GR. In particular, 
for $l=1$, one finds 
\beq
w_1''+\frac4r w_1'=0\,,
\eeq
whose  solution is of the form 
\begin{eqnarray}
w_1=-2\frac{J}{r^3},
\end{eqnarray}
where $J$ is an integration constant corresponding physically to the angular momentum. The moment of inertia is  defined as $I= J/\Omega$. 
The integration constant $J$ can then be matched to the interior solution for $w_1$  so that  the moment of inertia of the neutron star $I$ can be written in the integral form 
\begin{eqnarray}
I=\frac{(hf)^{\frac{1}{2}}}{6 F}\int_0^{R}\frac{h^{\frac{1}{2}}(P+\rho )r^4}{f^{\frac{1}{2}}}  \left(1+\frac{w_1}{\Omega}\right)\, dr\,.
\end{eqnarray}

As discussed in the introduction, interesting universal relations have been obtained in GR, in particular relating  the normalized moment of inertia 
\beq
\tilde I=I/(MR^2)
\eeq
 and the stellar compactness
\beq
{\cal C}=\frac{M}{R}\,,
\eeq
as well as the dimensionless moment of inertia 
\beq
\bar I=I/M^3
\eeq
 with the compactness ${\cal C}$.
These universal relations are respectively of the form:
\begin{eqnarray}
  \frac{I}{M R^2}&=&\ta_0+\ta_1\,  {\cal C}+\ta_4\,  {\cal C}^4\,,\label{URL}\\
  \frac{I}{M^3}&=&\ta_{-1}\, {\cal C}^{-1}+\ta_{-2}\,  {\cal C}^{-2}+\ta_{-3}\,  {\cal C}^{-3}+\ta_{-4}\,  {\cal C}^{-4},\label{URNL}
  \end{eqnarray}
  where the constants $\ta_i$ for $i=\{-4,-3,-2,-1,0,1,4\}$  can be estimated numerically.
  
\begin{figure}[h]
\centering
\includegraphics[scale=0.8]{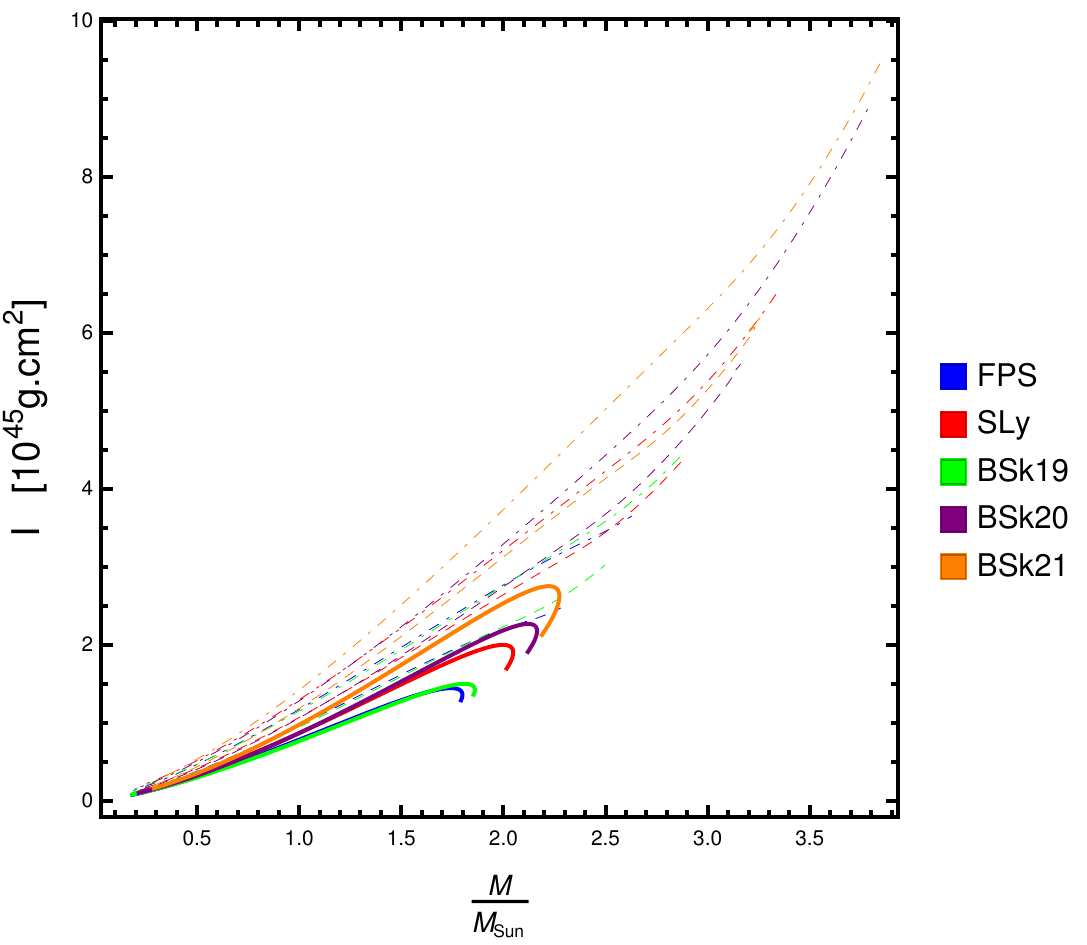}  
\caption{\small Relation between  the neutron star's moment of inertia and its mass $M$, using the same notations and parameters as in Fig.~\ref{mrs}.}
\label{IM}
\end{figure}

\begin{figure}[h]
\centering
\includegraphics[scale=0.8]{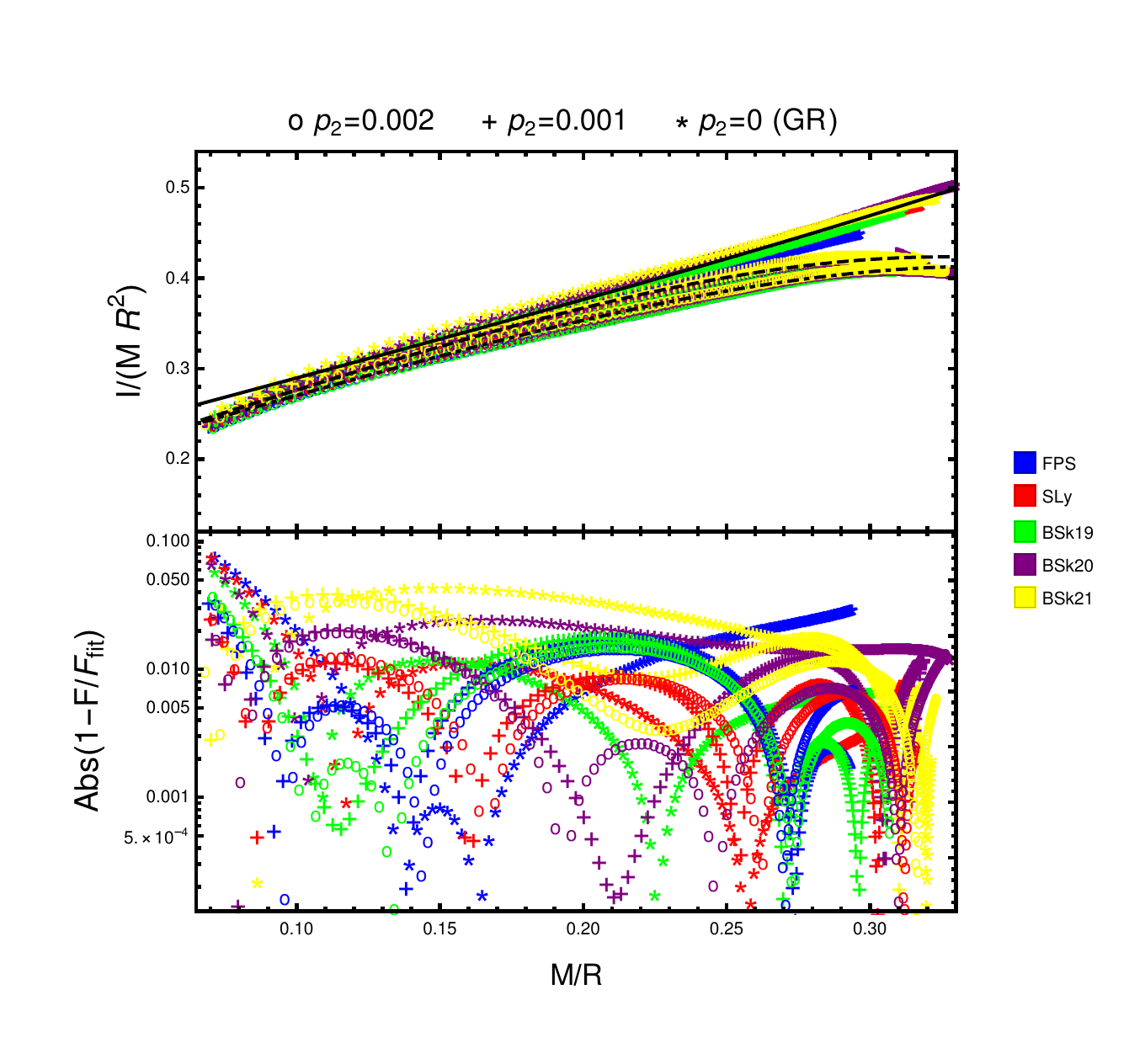}  
\caption{\small The variation of the normalized moment  of inertia $I/(MR^2)$ with respect to the compactness of the star $M/R$. We plot the fitted universal relation (\ref{URL}) for the cases $p=0$ (GR), $p=0.001$ and $p=0.002$, with continuous, dashed and dot-dashed lines, respectively.}
\label{imrs2}
\end{figure}

\begin{figure}[h]
\centering
\includegraphics[scale=0.8]{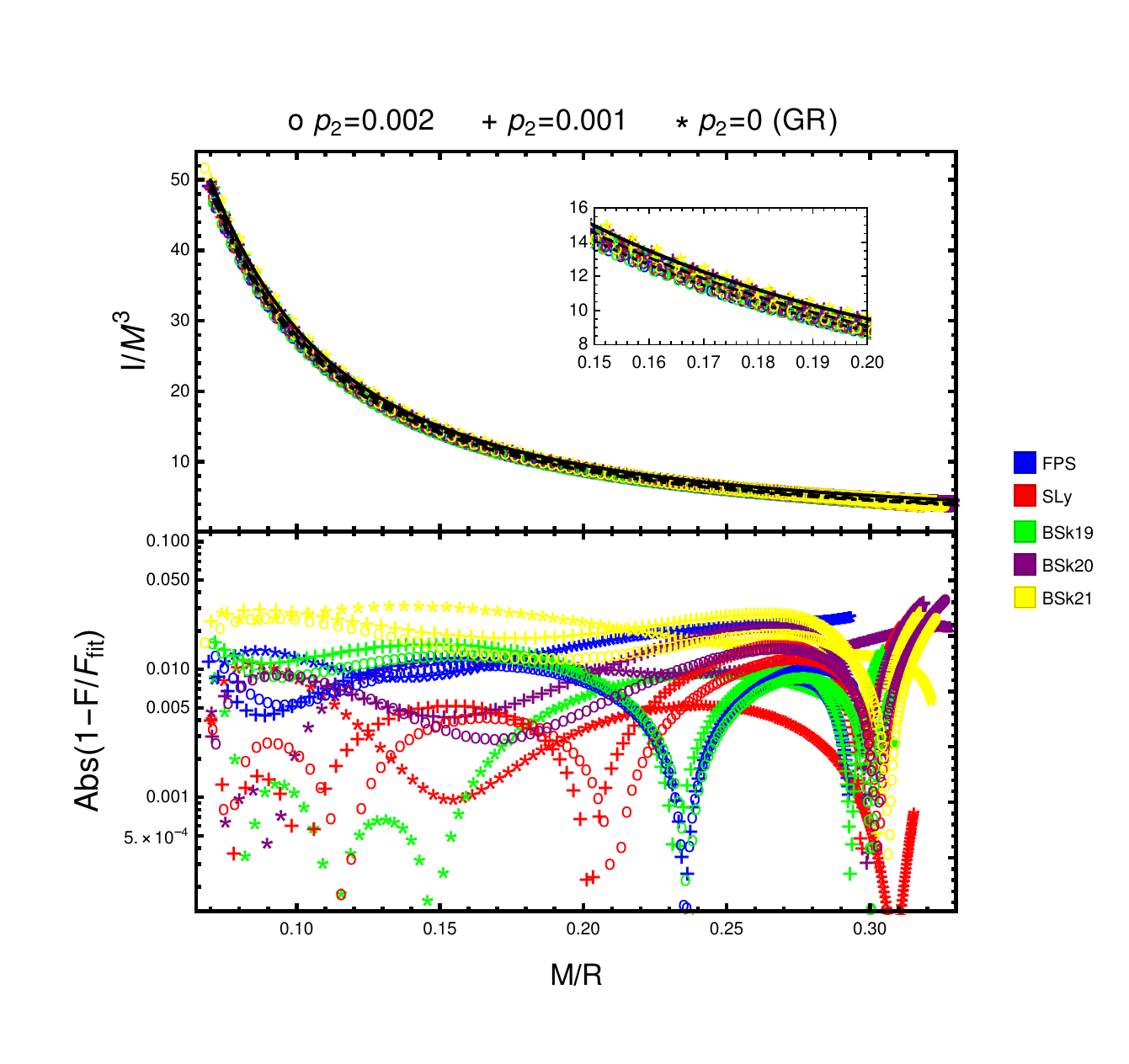} 
\caption{\small The variation of the normalized moment  of inertia  $I/M^3$ with respect to $M/R$. Using the result of Table \ref{T2},  we plot the  universal relation (\ref{URL}) for the cases: $p=0\,(GR)$, $p=0.001$ and $p=0.002$ with continuous, dashed  and dot-dashed lines, respectively. }
\label{im3}
\end{figure}

In Fig.~\ref{IM}, we have plotted the relation between the moment of inertia and the mass for the five equations of state.
 The values of $p$ that we have used in our numerical integrations are $p=\{0,\;0.001,\;0.002\}$. 
We then show, 
in Fig.~\ref{imrs2} and Fig.~\ref{im3},  that these results can be fitted by  universal relations (\ref{URL}) and (\ref{URNL}), both in GR and for  our modified gravity model, noting that the quantity $\mid 1-I/I_{fit}\mid$ is less than $10\%$ for the relation (\ref{URL}) and less than $5\%$ for the relation (\ref{URNL}).
We have listed the corresponding  coefficients  $a_i$  in Tables \ref{table1} and \ref{table2}.  Note that our  values in the GR case differ from those given in \cite{Lattimer:2004nj,Breu:2016ufb}  because we are using a  different  sample of equations of state, which illustrates that the universality of the above relations remains somewhat relative.

\begin{table}[ht]

\centering 
\begin{tabular}{|c|c|c|c|}\hline
            & GR &p=0.001 & p=0.002\\\hline
 $\ta_0$ & 0.205  $\pm$ 0.003 & 0.175  $\pm$ 0.002 & 0.179  $\pm$ 0.001\\\hline
 $\ta_1$ & 0.849  $\pm$ 0.017  & 1.018  $\pm$ 0.01 & 0.918  $\pm$ 0.008\\\hline
 $\ta_4$ & 1.23  $\pm$ 0.31 & -7.41  $\pm$ 0.19 & -5.85  $\pm$ 0.14\\\hline
$\text{I}_{\chi^2}$ & 41.86\; $10^{-6}$ & 13.42 \;$10^{-6}$ & 8.96\; $10^{-6}$\\ \hline
\end{tabular}
\caption{The coefficients of the universal relation (\ref{URL}).}\label{T1} 
\label{table1}
\end{table}

\begin{table}[ht]
\centering 
\begin{tabular}{|c|c|c|c|}\hline
            & GR &p=0.001 & p=0.002\\\hline
 $\ta_{-1}$ & 0.906  $\pm$ 0.028& 0.410 $\pm$ 0.028& 0.419  $\pm$ 0.023  \\\hline
 $\ta_{-2}$ & 0.184  $\pm$ 0.013& 0.342  $\pm$ 0.012 & 0.317  $\pm$ 0.010\\\hline
 $\ta_{-3}$ & 0.005  $\pm$ 0.0017& -0.0135  $\pm$ 0.0016   & -0.0111  $\pm$ 0.0013\\\hline
 $\ta_{-4}$ & -0.00036  $\pm$ 0.00007 & 0.00031  $\pm$ 0.00006 & 0.00024  $\pm$ 0.00005 \\\hline
$\text{I}_{\chi^2}$  & 0.021 & 0.022 & 0.015 \\\hline
\end{tabular}
\caption{The coefficients of the universal relation (\ref{URNL}).}\label{T2} 
\label{table2}
\end{table}

\section{Conclusions}
In this work, we have explored neutron stars within a simple class of modified gravity models, based on a single-parameter subfamily of DHOST theories.  In this gravity models, one recovers the usual Schwarzschild solution outside a spherically symmetric and static matter distribution, even if the scalar field profile is nontrivial outside. By contrast, in the interior of the star, the radial profiles for the geometry and for the matter deviate from the GR  situation.

To quantify the deviations more precisely, we have used several equations of state to describe the equation of state of the neutron star matter. In each case, for different values of the modified gravity parameter, we have computed numerically the neutron star profile for a range of values for the central energy density. In comparison with the GR values, we have found that neutron stars with significantly higher masses and radii are possible.

Intuitively, this different behaviour can be understood by attributing to the modified gravity effects an effective energy density and pressure. Interestingly, the equation of state for this effective matter is typically such that the ratio $P_\varphi/ \rho_\varphi$ varies between $-1/2$ and $-1/3$,  according to Fig.  \ref{rhoeff}. And the effective energy density is negative in the inner layers of the star before becoming positive in the outer layers.
In view of this surprising property, an important task, left for the future, to test the viability of these neutron stars models would be  the study of their stability with respect to radial or non-radial perturbations. 

In order to further characterise the phenomenological difference between these neutron stars and their GR counterparts, we have also investigated the status of the relations between the moment of inertia and the compactness of the star. The interest of such universal, or quasi-universal,  relation is that it is weakly sensitive on the equation of state, so that one can evade the obstacle of the large uncertainty on the equation of state for neutron star matter. Therefore a precise measurement of these relations would in principle enable us  to discriminate between different gravity theories.

Finally, let us note that our study has been restricted to a simple one-parameter subfamily of DHOST theories. It would be interesting to explore other sectors of this large family of modified gravity theories and see whether they lead to deviations of GR that can be distinguished qualitatively and quantitatively.

\newpage

\appendix
\section{Coefficients of the first-order system}
As discussed in the main text, we rewrite the equations of motion in the matricial form
\beq
A\, \frac{d Y}{dr}=B\,, \qquad Y\equiv (f,h, \psi')^T\,,
\eeq
where the first line corresponds to (\ref{ef}) and the second line to (\ref{eh}). The last line is obtained from the radial derivative of Eq.(\ref{ex}), which
can be written formally as
\beq
\psi'^2 =\Lambda(f, h, \rho, P,r)\,.
\eeq
One obtains
\beq
\frac{\partial \Lambda}{\partial f} f'+\frac{\partial \Lambda}{\partial h} h'+\frac{\partial \Lambda}{\partial \rho} \rho'+\frac{\partial \Lambda}{\partial P} P'+\frac{\partial \Lambda}{\partial r}  -2\psi'\psi''=0\,,
\eeq
and after rewriting  $P'$ and $\rho'$ in terms of  $f'$ by using (\ref{e4}) and (\ref{sound}), one can  identify  the corresponding coefficients in the matrices $A$ and  $B$.

Finally, the coefficients of the matrices $A$ and $B$ in (\ref{system}) are given by
\begin{eqnarray}
&A_{11}=\frac{r }{2 f^2}\left(\frac{8 h p^2}{h p-f \psi '^2}+\frac{6 f \psi'^2}{h}-f \kappa -6 p\right)\,, \quad A_{12}=0\,, \quad A_{13}=\frac{4 p r \psi '}{h p-f \psi'^2}\\
&A_{21}=0\,, \quad A_{22}=-\frac{f}{r}A_{11}\,,\quad A_{23}=4\psi'+\frac{h}{r}A_{13}\\
&A_{31}=\partial_f\Lambda-\frac{P+\rho}{2f}(\partial_P\Lambda+\partial_\rho\Lambda/c_m^2)\,, \quad A_{32}=\partial_h\Lambda,\qquad A_{33}=-2\psi',\\
&B_1=-\frac{p }{f}\left(\frac{4 p h}{f \psi '^2-p h}+h+3\right)+\frac{1}{2} (1-h) \kappa -\frac{1}{2} h P r^2-(h+3)\frac{\psi'^2}{h}\\
&B_2=-\frac{1}{2r} h^2 \left(2 \kappa +P r^2-\rho  r^2\right)-\frac{4 \psi '^2}{r}+\kappa \frac{h}{r} -\frac{h}{r}B_1,\qquad B_3=-\partial_r\Lambda\,,
\end{eqnarray}
where $\psi'$ can be expressed in terms of the other quantities following  (\ref{ex}).

\section{Behaviour near the center of the star:}
In order to determine the initial conditions at the center of the star,  we expand the metric functions,  pressure and energy density density in the form
\begin{eqnarray}
\label{expansions}
h=1+h_2 r^2,\qquad f=f_c+f_2 r^2,\qquad P=P_c+P_2 r^2 \quad {\rm and} \quad \rho=\rho_c+\rho_2 r^2
\end{eqnarray}
 where the coefficient $h_2$, $f_c$, $f_2$, $P_c$, $P_2$, $\rho_c$ and $\rho_2$ are  constants\footnote{The subscript 'c' stands for 'central' and denotes  the value of the  function at $r=0$.}. Note that $h(r=0)=1$ since the spatial geometry must be locally Euclidean at $r=0$ (otherwise there would be a conical singularity). Similarly, one must have $h'(r=0)=f'(r=0)=P'(r=0)=0$  to satisfy regularity conditions.  
 
By  substituting  the above expansions in the expression (\ref{ex}) for the scalar field, one determines the expansion of $\psi'$ near $r=0$.  Then, substituting all these expansions into the two other equations of motion, one obtains two relations that enable us to find  $f_2$  and $h_2$ in terms of the other  coefficients. $P_2$ can then be obtained from (\ref{e4}).
Eventually,  we find that  the scalar field,  metric  and matter behave near the center as 
\begin{eqnarray}
\psi'^2&=&\frac{ \left(f_c \kappa  \left(6 P_c+5 \rho_c\right)-2 p \left(6 P_c+ \rho_c\right)\right)}{6  \left(f_c \kappa +2 p\right) \left(f_c \kappa +6 p\right)}p r^2,\\
f&=&f_c+f_c\frac{ \left(2 f_c p \left(15 P_c+7 \rho_c\right)+f_c^2 \kappa  \left(3 P_c+\rho_c\right)\right)}{6 \left(f_c \kappa +2 p\right) \left(f_c \kappa +6 p\right)}r^2,\\
h&=&1+\frac{f_c^2 \kappa  \rho_c-f_c p \left(24 P_c+10 \rho_c\right)}{3 \left(f_c \kappa +2 p\right) \left(f_c \kappa +6 p\right)}r^2,\\
P&=&P_c-\frac{\left(P_c+\rho_c\right) \left(2 f_c p \left(15 P_c+7 \rho_c\right)+f_c^2 \kappa  \left(3 P_c+\rho_c\right)\right)}{12 \left(f_c \kappa +2 p\right) \left(f_c \kappa +6 p\right)}r^2\,.
\end{eqnarray}
These expansions are used as initial conditions to solve the radial differential equations  numerically.

Note that the first relation implies the following constraint on $p$: \begin{eqnarray}
p<\frac{f_c \kappa  \left(6 P_c+5 \rho_c\right)}{2  \left(6 P_c+ \rho_c\right)}.
\end{eqnarray} 
This  condition automatically implies  $P''(0)<0$, which is necessary to get a  physically viable profile for the pressure. 

Substituting the above expansions into the expressions for the effective energy density and pressure, one finds that they are given at the center of the star by the values
\begin{eqnarray}
\rho_{\varphi,c}  &=&-\frac{6 p(f_c \kappa  (4 P_c+3 \rho_c)+2 p \rho_c)}{(f_c \kappa +2 p) (f_c \kappa +6p)},\\
P_{\varphi,c}  &=&\frac{2 p (4  f_c \kappa  \rho_c+5  f_c \kappa  P_c-6 P_c p)}{( f_c \kappa +2 p) ( f_c \kappa +6 p)}\,.
\end{eqnarray}

Finally, let us mention that a similar analysis for  the non-rotating case leads to 
\beq
w_1=w_{1,c}+\frac{f_c \left(P_c+\rho_c\right) (w_c+\Omega )}{5 \left(f_c \kappa +2 p\right)}r^2\,,
\eeq

which is useful to integrate (\ref{eq_w_l}) for $l=1$.

\newpage

\bibliographystyle{utphys}
\bibliography{DHOST_NS_biblio}

\end{document}